# The role of sign in students' modeling of scalar equations

Kate Hayes[*], Michael C. Wittmann[*§†]
University of Maine, Orono ME 04469-5709
* Center for Science and Mathematics Education Research
§ Department of Physics and Astronomy
† College of Education and Human Development
kate.mccann@umit.maine.edu, wittmann@umit.maine.edu

## INTRODUCTION

Helping students set up equations is one of the major goals of teaching a course in physics that contains elements of problem solving. Students must take the stories we present, interpret them, and turn them into physics; from there, they must turn that physical, idealized story into mathematics. How they do so, and what problems lie along the way, are a major source of difficulty for us, as instructors. In this paper, we consider just one such difficulty, getting the plus and minus signs correct when setting a net force equal to mass times acceleration. Even in such simple equations, we find that students make common errors in how they connect the mathematics and the physics. Specifically, we have seen college physics students use physical and mathematical reasoning inconsistently when determining signs of terms in equations. The problem seems to lie in how a vector equation gets interpreted into a scalar equation (whose form depends on one's choice of coordinate system).

Our work takes place in a course using the *Intermediate Mechanics Tutorials* [1,2] in a sophomore level course in mechanics (using Newtonian and only a little Lagrangian mechanics), but the results are common to other forms of instruction and relevant to instruction at all levels of physics. Many of our results come from informal observations, but we present here results from students videotaped during instruction while using small-group activities developed in the mindset of existing tutorial instruction. [3-5]

## WHEN UNEXPECTED PARAMETERS AFFECT THE FORMALISM

### Problem Statement

As part of a unit on air resistance, students had to answer the following question:

"A ball is thrown vertically downward at greater than terminal velocity from the top of a tall building. It experiences an air resistance force proportional to $v$. Find an equation that describes the velocity of the ball with respect to time. Let $+y$ be in the downward direction."

Note two uncommon elements of this problem. First, the ball is thrown at greater than terminal velocity. (We have found that many students do not expect it to slow down to terminal velocity.). Second, we explicitly define the coordinate system because we have found that certain coordinate system choices make some parts of the problem easier and others harder for students. We were hoping to raise a specific issue about signs with students, as is shown below.

Using Newton's Second Law, the ball's velocity as a function of time can be written in vector form (eqn. 1a) and scalar form (eqn. 1b). as

$$m\mathbf{a} = m\mathbf{g} - c\mathbf{v} \qquad (1a)$$

$$ma = mg - cv \qquad (1b)$$

Note that in equation 1b the signs of the "$ma$" and "$cv$" terms do not depend on one's choice of coordinate system, while the sign of the "$mg$" term does. Equation 1b gives the acceleration of the object as determined by the downward weight (positive, since it is in the direction of $+y$) and the upward air resistance term (negative, regardless of direction of force, since the value of $v$ may have negative or positive values depending on the direction of motion, and the force always points opposite the direction of motion).

From this point forward, one has many possible pathways for further solutions. One can convert $a$ to a differential term, $dv/dt$, and solve the differential equation shown in equation 1c through separation of variables, etc. Though this was expected of students, we do not describe this part of their work in this paper any further.

$$m\, dv/dt = mg - cv \quad (1c)$$

Instead, we focus on a common problem in setting up the equation shown in 1b. At one point in the tutorial, six of the 11 students in the class had written the equation shown in eqn. 2, with a minus sign used incorrectly:

$$-ma = mg - cv \quad (2)$$

How did they arrive at this equation? Had there been a minus sign before the "mg" term, they might only have mistaken the direction of the coordinate system. Had there been a plus in front of the "cv" term, they might have made the same error. Why, though, was there a minus sign incorrectly placed in front of the "ma" term? Answering this question reveals much about how students use physics to guide their thinking about mathematics, and how dynamic their thinking is while setting up equations. We note that most of the students incorrectly writing eqn. 2 modified their equation to use the correct one after either group discussion or discourse with the instructor (MCW). We present a single student's reasoning in detail because he was most articulate about the common problem we observed.

## Context-Dependent Reasoning and the Conflation of Ideas

One student, "Max," strongly argued to keep the minus sign in front of the acceleration term and his argument reveals the significant difficulty students have in representing a physical situation with mathematics. Unfortunately, Max was not videotaped as part of our effort to videotape classroom interactions (but for a snippet heard in the background, consistent with the interview results shown below); therefore, one author (KM) invited him to discuss his ideas further outside of class. The interview with Max lasted for approximately 15 minutes. There were no set procedures or questions for the interview. Max was asked to explain his reasoning during the tutorial and the interviewer asked any questions to clarify points and rationale. A follow-up interview was done the next week.

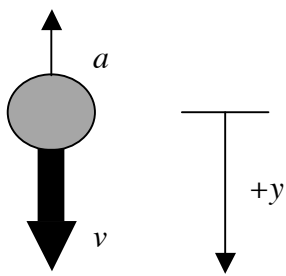

**FIGURE 1.** Max's motion diagram of a ball falling while moving at greater than terminal velocity, indicating acceleration and velocity and a coordinate system.

During the first interview, Max sketched a diagram indicating velocity and acceleration of the ball (recreated in Figure 1). Max argued that the minus in front of the acceleration term (as written in eqn. 2) was necessary to prevent Newton's second law (as written in eqn. 1b) from leading to physically incorrect solutions.

Max considered two different ways to manipulate Newton's Second Law. In the relatively standard method, $v$ is used to solve for $a$, so that eqn. 1b indicates that "$mg - cv$" *leads to* an acceleration, $a$. Max knows that the force of air resistance is greater than the force of gravity on the ball and therefore, in the given coordinate system, that the quantity ($mg - cv$) is negative. The "$ma$" term is negative, therefore, making the acceleration, $a$, negatively valued. This agrees with his understanding of the coordinate system, in which positive $y$ is in the downward direction: the acceleration of the ball is pointed up, which is in the negative direction. We do not give transcript of this reasoning, since it is relatively standard.

But, when asked to solve the problem as stated (namely, to find the velocity, $v$ as a function of time), Max reasons differently. In this solution, he asserts, one uses the acceleration term, $a$, to solve for the velocity term, $v$. This leads to results inconsistent with his understanding of the coordinate system. Several "errors" cropped up at this point; we wish to highlight his assumptions (and their consequences). We give an extensive transcript to illustrate different pieces of his reasoning.

In order to solve for velocity given an acceleration, Max stated several times that acceleration is assumed to be positive and repeated this assertion several times during the interview. We are not sure why exactly he assumed this, perhaps because there is no negative sign outside of the "$ma$" quantity. (Note that transcript lines are numbered in the order they were spoken, though some lines are skipped.)

1  Max: See, this equation right here, if you
2  use *a* to solve for *v*, it assumes *a* is positive.
3  There's no negative in there. It assumes *a* is
4  positive. So -- *ma* minus *mg*, but we know
5  *mg* is smaller than *cv*, that's [*ma* – *mg*] still a
6  positive number. So, *a* assumed to be
7  positive, divide by negative *c*, *v* is a negative
8  number. And that, if you solve it backwards
9  like that, it just said *a* is going down and *v* is
10 going up.

Max was adept at algebraic manipulation and employed the admirable technique of checking his response against the givens in the system. (From a behavioral standpoint, he is engaged in very valuable activities, which we try to promote in tutorial instruction.) Max's conflict arose when checking answers against expectations (lines 8 to 10). Earlier, he assumed that *a* is a positive number (lines 2 to 4, 6 to 7, and interpreted in line 9). In addition, he assumed "m*a* – mg" was positive (lines 7–8). We do not know why he made this statement, but, having done so, the rest of his reasoning is clear. Given a positive value of *a* and "m*a* – mg"*,* Max arrived at the conclusion that *v* (which he knows is downward, and defined as positive) must instead be negative, therefore up.

Max proposed a solution – that he manipulate the equation to arrive at the correct situation.

```
11  M: I determined that in order to solve in that
12     manner, to go from a to solve for v, to get it
13     in the correct frame, you needed a minus
14     sign in front of ma.
15  Interviewer: Alright.
16  M: 'Cause that way if you have negative ma
17     equals mg minus cv, you go through that
18     same process of subtract mg, mg we know is
19     less than that [ma], so it's still gonna be
20     negative out here. Then you divide by c or
21     negative c, that becomes a positive number,
22     which makes v positive, which is going
23     down. And a was initially negative number,
24     which is how it should be.
```

In lines 11 to 14, Max clearly manipulates the given equation in order to have one that leads to a solution (positive *v*) consistent with the problem. He writes out eqn. 2 as he reasons through the problem again. In lines 16 to 24 he repeats his reasoning from before, pointing at the different parts of eqn. 2. Note that he again makes an assumption about the size of the "m*a*" term compared to the "mg" term. Finally, note his conflation of "a minus sign in front of m*a*" in lines 13 to 14 and "*a* was initially negative number," two statements that are not the same at all (though he says, "which is how it should be" in line 24, as if they are the same).

We have come up with two terms to describe Max's reasoning. When writing "–m*a*" he places a minus sign in front of the term; we call this the *outer minus*, for obvious reasons. At the same time, it is possible that a quantity have negative values, even when written with a letter that has no negative sign out front. We call this the *inner minus*. The *inner minus* is what distinguishes a constant (assumed positive in physics) from a variable (which can be positive or negative). At times, Max treats variables as if they have the properties of constants.

Later in the interview, Max was asked again about the acceleration term and again combined these two ideas:

```
25  I: What if we just said a was negative?
26  M: If you say a is negative, it comes out
27     properly…makes acceleration negative,
28     which is in the up direction, velocity goes in
29     the down direction...
30  I: Uh huh.
31  M: ...which is how it should be. That was
32     my reasoning on it.
```

Max (in lines 26–29) seems to accept the idea of using a negative value for *a*. He has worked through the algebraic reasoning and properly evaluates his answers and their physical meaning. In lines 31–32, Max states the consistency of his use of the *outer minus* in eqn. 2 with the use of the *inner minus* suggested by the interviewer in line 25. As a consequence, he writes an equation that violates Newton's Second Law by inserting his choices of sign into the formalism.

In the later follow-up interview (not discussed in more detail in this paper), Max is questioned about his view on sign and direction as defined by vector quantities translated into scalars in a given coordinate system. Max states at one point, "It really doesn't have any sign until I assign it one." This seems to be the crux of the matter: *outer minus* and *inner minus* are equivalent when he decides on the sign. He feels he is consistent in his actions and words.

## SUMMARIZING STUDENT REASONING

We observe several valuable elements to Max's reasoning, while also observing a fundamental inconsistency in the way he approaches signs in Newton's Second Law. We wish to make three points about his reasoning, the first about excellent habits of mind, the second about his inconsistent use of variables, and the third about how we, as researchers and instructors, should avoid thinking of his work in terms of robust and consistent errors.

### Seeking Consistency in Math and Physics

Max admirably checks his mathematical responses against his physical intuition, carefully interpreting the results of his work in ways that are not necessarily common even for sophomore physics majors. Max is dynamic in how he manipulates the mathematical equation. In lines 5–6 and 18–19, he uses terms (like "m*a*" or "mg") as chunks or tidbits that he can move about and manipulate. He distinguishes between these terms and the variables he's interested in, as seen in lines 7 and 20–21, where he divides by the coefficient "c" to arrive at a value for "*v*". In lines 9–10, 22–24, and 27–29, Max explicitly connects the coordinate system, algebraic equation, and physical situation and seeks consistency between the different descriptions of the system.

## Letters as Constants or Variables

Max sees no difference in the use of *outer minus* and *inner minus* and asserts that the conflation of *outer* and *inner minus* is consistent with the physics (see lines 31–32). Perhaps the problem lies in his asserting that letters are like constants (which have a value that is only positive). He explicitly states (lines 3–4 and 6–7) that the acceleration is assumed to be positive and implements this assumption as shown in equation 2. More to the point (and a source of the name *inner minus*), he asserts in line 3 "there's no negative in there" while pointing at the letter $a$, the acceleration.

At this point, Max's use of consistency breaks down. As shown in lines 25 to 32, he does not see the inconsistency of having acceleration, $a$, be possibly negative while also assigning it a minus sign, as $-a$ (with $a$ assumed positive).

## Context-Dependent Activation

Most of Max's reasoning is not problematic. His use of consistency is commendable, and he can work well with equations. Problems arise when he combines *outer minus* and *inner minus* and arrives at a false statement of Newton's Second Law. If **F** = –m**a** *or* m**a**, depending on one's needs in the problem, then the equation depends on the local choices of the problem solver and not the universality and consistency of the mathematics and the physics. It is this breakdown which worries us most, as instructors.

Max's assignment of positive values to letters seems not to be arbitrary - he does it with both velocity and acceleration, both algebraically and numerically. His assumption seems to be that *he* can choose a value for a variable. Thus, we see Max creating his own values for quantities and assuming positive values when working numerically, and we see him asserting positive values for variables when the minus sign is not visible.

When solving for $a$, even though Max knows it must be negative, he does not insert a minus sign by hand, but instead accepts finding its *inner minus* as he solves for $a$. By contrast, when using $a$ to solve for something else, he feels the need to insert the (*outer*) minus by hand instead of letting the *inner minus* exist. We cannot capture his reasoning by attributing to him a global "misconception" about the properties of variables vs. constants or about outer vs. inner minus. His conflation is context dependent but nonetheless predictable, appearing when the variable in question is used to solve for something else.

This shouldn't be surprising: When treating a variable as a "given" to solve for something else, it's easy not to think about the distinction between given constants and "given" variables. In the complicated task of connecting algebraic formalism, geometric concerns as represented by coordinate systems, and physically meaningful descriptions, many different ideas play a role. These might or might not be used consistently. Such lack of consistency can be described using other language, of knowledge-in-pieces or resources [6–9].

## DISCUSSION

We have used an example from a single student's responses to a common physics question to describe in detail how many students in our course (and in other settings, as we have found informally) reason about the equations they use to model the physics. Our student, Max, interprets the physical meaning of coordinate systems very carefully and systematically, but makes assumptions about variables that lead to inconsistent definitions of Newton's Second Law.

Max's reasoning about the physics problem raises fundamental questions about how students think about variables. For Max, the type of problem determines whether variables can have negative or only positive values, raising the issue of agency in determining the necessary signs. Though we have only looked at one student in detail, we find that the problems of assumed positive values and agency in determining signs are more common than we wish.

Though this paper is not about addressing the types of reasoning shown by Max, we suggest in short that effective instruction on this issue would address Max's use of *outer* and *inner minus* possibly by emphasizing the use of consistency and helping students develop the idea that one cannot choose signs arbitrarily when working with variables.